\documentclass[prl,twocolumn,amsmath,amssymb,citeautoscript,showkeys,floatfix,superscriptaddress]{revtex4-2}

\usepackage{chemformula} 
\usepackage[T1]{fontenc} 
\usepackage{siunitx}
\usepackage{amssymb}
\usepackage{amsmath}
\usepackage{graphicx}
\usepackage{multirow,makecell}
\usepackage{booktabs}
\usepackage{footnote}
\usepackage{comment}

\usepackage{hyperref}
\hypersetup{
    colorlinks,%
    citecolor=blue,%
    linkcolor=blue,%
    urlcolor=blue
}

\providecommand{\U}[1]{\protect\rule{.1in}{.1in}}

\date{\today}

\begin{document}

\title{Direct evidence for carbon incorporation on the nitrogen site in AlN}

\author{Darshana Wickramaratne}
\affiliation{Center for Computational Materials Science, US Naval Research Laboratory, Washington,
D.C. 20375, USA}
\author{Mackenzie Siford}
\affiliation{Department of Physics, University of Alabama at Birmingham, Birmingham, Alabama 35233, USA}
\author{Md Shafiqul Islam Mollik}
\affiliation{Department of Physics, University of Alabama at Birmingham, Birmingham, Alabama 35233, USA}
\author{John L. Lyons}
\affiliation{Center for Computational Materials Science, US Naval Research Laboratory, Washington,
D.C. 20375, USA}
\author{M.E. Zvanut}
\affiliation{Department of Physics, University of Alabama at Birmingham, Birmingham, Alabama 35233, USA}
\begin{abstract}
We use photo-electron paramagnetic resonance (EPR) measurements and first-principles calculations to
identify and explain the properties of carbon in AlN.
We present clear evidence for carbon substitution on the nitrogen site (C$_{\rm N}$). We also clarify the origin of a widely
observed EPR spectra in AlN that, although often attributed to a deep donor defect, we demonstrate is surprisingly due to
C$_{\rm N}$. Finally, we show the presence of C$_{\rm N}$ is consistent with the absorption spectra at 4.7 eV observed in AlN.
\end{abstract}

\maketitle
Aluminum nitride (AlN) is one of the most promising ultra-wide band gap semiconductors
for applications in power electronics \cite{tsao2018ultrawide,lyons2024dopants}
and optoelectronics that operate in the deep ultraviolet \cite{kneissl2019emergence}.
Continued progress in defect spectroscopy and identification have shown that AlN exhibits a rich array of
defects \cite{soltamov2010identification,mason1999optically,soltamov2011identification,evans2006electron,son2011defects,
yan2014origins,schulz2011ultraviolet,koppe2016overview}.
Within this catalog of defect levels there has been considerable effort to identify the microscopic origin of
an absorption band at 265 nm (4.7 eV), which can potentially hamper the operation of AlN light
emitters that operate in the deep ultraviolet.
Carbon has been invoked as the origin
of this absorption signal \cite{gamov2021photochromism, collazo2012origin,alden2018point, irmscher2013identification}.  
However there is some debate as to whether carbon incorporates by substituting on the nitrogen site,
C$_{\rm N}$ \cite{alden2018point}, substituting on the Al site, C$_{\rm Al}$ \cite{gamov2021photochromism}, 
or as a complex \cite{irmscher2013identification}.  The question of
whether carbon alone provides a sufficient explanation for this absorption band has also been raised \cite{peters2023combination}.

Defect identification based on optical spectroscopy alone is challenging when there are multiple overlapping
optical bands \cite{yan2014origins, schulz2011ultraviolet}.
Spectroscopy techniques such as electron-paramagnetic resonance (EPR) provide a more precise fingerprint of
defects and can greatly expand our understanding of the microscopic origin of defects.
Indeed, several EPR studies, including our own, have identified
a spectrum with the following hyperfine (HF) parameters - $A_{\parallel}$ $\sim$ 3.9 mT and $A_{\perp}$ $\sim$ 1.9 mT
\cite{soltamov2010identification,soltamov2011identification,mason1999optically,evans2006electron}, where
$A_{\parallel}$ is the HF interaction parallel to the $c$-axis of AlN and $A_{\perp}$ is the HF interaction determined
perpendicular to the $c$-axis.
The defect attributed to this spectrum (which we henceforth refer to as D5, following Ref.~\cite{mason1999optically})
was initially posited to be a deep donor with the following suggested candidate defects, 
an Al interstitial (Al$_i$), a nitrogen vacancy ($V_{\rm N}$),
or oxygen substituting on the nitrogen site (O$_{\rm N}$).
The suggestion that the D5 defect might be a deep donor was based in part on 
electron nuclear double resonance (ENDOR) measurements \cite{evans2006electron}, which revealed the
hyperfine interaction of D5 arises
from an $S$=1/2 defect interacting with a nuclear spin of $I$=5/2, which was assumed to be the Al atoms of AlN.
At first glance, these studies do not appear to offer any insight into the microscopic origin of the absorption
band at 4.7 eV and its potential connection with the presence of carbon in AlN.

However, there are some issues that contradict the accepted wisdom that the D5 defect is a deep donor.
Previous DFT calculations of HF parameters ruled out Al$_i$
as the origin of D5 \cite{gerstmann2001transition}.
Follow-on EPR measurements together with first-principles calculations \cite{son2011defects}
cast doubt on D5 being due to $V_{\rm N}$.  Since first-principles calculations predict O$_{\rm N}$ is a $DX$ center \cite{Gordon14},
it is unlikely that D5 is due to O$_{\rm N}$, since the EPR active charge state (O$_{\rm N}^{0}$) is metastable, while D5 is observed in 
as-grown AlN.

In this Letter, we show these two puzzles in AlN, the role of carbon in causing the 4.7 eV absorption signal and the
unknown source of the D5 EPR signal, are in fact closely linked. We present a unified physical picture showing that the D5 defect is
C$_{\rm N}$, which is a deep acceptor in AlN. Measuring the EPR activity as a function of photon energy leads to two clear transitions
that either quench or excite the EPR signal.  Our first-principles calculations indicate that the hyperfine parameters and the optical transitions
are explained by the C$_{\rm N}$ (0/$-$) acceptor level. Our study corrects a long-standing misconception that the D5 defect in AlN is a deep donor, and provides clear evidence for the incorporation of C$_{\rm N}$ in AlN.

We performed our photo-EPR measurements on AlN substrates from Hexatech and Crystal IS,
both of which were grown via physical vapor transport.  The 7 mm $\times$ 7 mm $\times$ 0.5 mm AlN substrate
from Hexatech, which has surface coloration changing from amber to clear, was cut into 2.3 mm wide pieces,
and labeled sample A, sample B and sample C.  Several similarly sized pieces, collectively called sample D,
were cut from an AlN wafer grown by Crystal IS.  In Table \ref{tab:sims} we list the concentration of
common impurities that were measured by secondary ion mass spectrometry (SIMS)
to a depth of 6 um on sample D and sample A, 
a piece with color and EPR response similar to sample B.

\begin{table}[]
\begin{center}
\caption{Secondary ion mass spectrometry (SIMS) measurements
of the concentration of common impurities Si, O, and C in sample A and sample D}
\setlength{\tabcolsep}{6pt} 
\renewcommand{\arraystretch}{1.4} 
\begin{tabular}{l|cc}
  \toprule\toprule
  {Impurity} & {Sample A (cm$^{-3}$)} & {Sample D (cm$^{-3}$)} \\
  \midrule
  Si & 2$\times$ 10$^{18}$ & 2$\times$ 10$^{17}$ \\
  O & 3$\times$ 10$^{18}$ & 3$\times$ 10$^{17}$ \\
  C & 1 $\times$ 10$^{19}$ & 2$\times$ 10$^{17}$ \\
   \bottomrule\bottomrule
\end{tabular}
\label{tab:sims}
\end{center}
\end{table}

10 GHz EPR measurements were obtained from each AlN sample between 300 K and
4 K with the magnetic field parallel to the $c$-axis.
Additional room temperature measurements were obtained from sample B
with the magnetic field rotated in the a-plane.
The spectra at different orientations were simulated using EasySpin \cite{stoll2006easyspin}
to determine the $g$-factor and hyperfine values.

Photo-EPR was carried out on samples C and D using light-emitting diodes (LEDs)
with the magnetic field along the $c$-axis. For the photo-quenching experiments,
first, a 265 nm LED (4.68 eV) with 0.5 mW power illuminated the sample through
the slits of the EPR cavity to generate the D5 signal. Then, the sample
was illuminated with LEDs of selected photon energy from 1.2 eV to 3.4 eV,
using neutral density filters to maintain a constant photon flux.
Between the measurements with each LED, the EPR signal was regenerated using the 265 nm LED.
Since the shape of the EPR signal did not change as the sample was illuminated,
a comparison of the EPR amplitude at a selected wavelength with that obtained
using the 265 nm LED was used to determine the relative number of spins measured at each wavelength.
The error in the relative amplitude is $\pm$0.1.  A similar procedure was followed for
photo-excitation of the EPR signal, except the signal always started near zero amplitude.  LEDs with peak photon energy between 
3.22 eV and 4.68 eV 
were used to increase the EPR signal, and a 470 nm (2.64 eV) LED
was used to quench the signal after each excitation. 
The number of spins generated
after illumination with the 265 nm LED for 30 minutes was 8$\times$10$^{14}$ for sample C and 2$\times$10$^{14}$
for sample D.
The average spin density, obtained by dividing by the sample volume,
was 1$\times$10$^{17}$ cm$^{-3}$ in sample C and 1.4$\times$10$^{16}$ cm$^{-3}$ in sample D.

To interpret our photo-EPR measurements we also perform first-principles calculations
based on density functional theory (DFT) \cite{HK64,KohnPR1965}, using the projector-augmented wave (PAW)
potentials \cite{BlochlPRB1994}
as implemented in the Vienna Ab-initio Simulation Package (VASP) \cite{KressePRB1996,kresse1994ab}.
We use the Heyd-Scuseria-Ernzerhof (HSE) hybrid functional \cite{HeydJCP2003,*HeydJCP2006}
for all of our calculations.  The energy cutoff for the plane-wave basis set is 500 eV.
The fraction of nonlocal Hartree-Fock exchange is set to 0.33 for AlN; this results in band gaps
and lattice parameters that agree with the experimental values.
Defect formation energies and thermodynamic transition levels are calculated using the standard supercell
approach \cite{FreysoldtRMP2014} with 96-atom and 288-atom supercells.
The lattice parameters of the supercell are held fixed while the atomic coordinates are relaxed
until the forces are below 5 meV/Angstrom.  The formation energies and charge-state transition levels of defects
are determined using a single ($\frac{1}{4}$,$\frac{1}{4}$,$\frac{1}{4}$) $k$-point.
Spin polarization is included for all of our calculations.
Optical absorption energies are determined through the construction of a one-dimensional configuration coordinate
diagram using the Nonrad code using the 96 atom supercell calculations \cite{turiansky2021nonrad}.
Hyperfine parameters for the paramagnetic charge state of each defect were calculated using a
288-atom supercell with a (2$\times$2$\times$2) $\Gamma$-centered $k$-point grid.

\begin{figure}[!h]
\includegraphics[width=8.5cm]{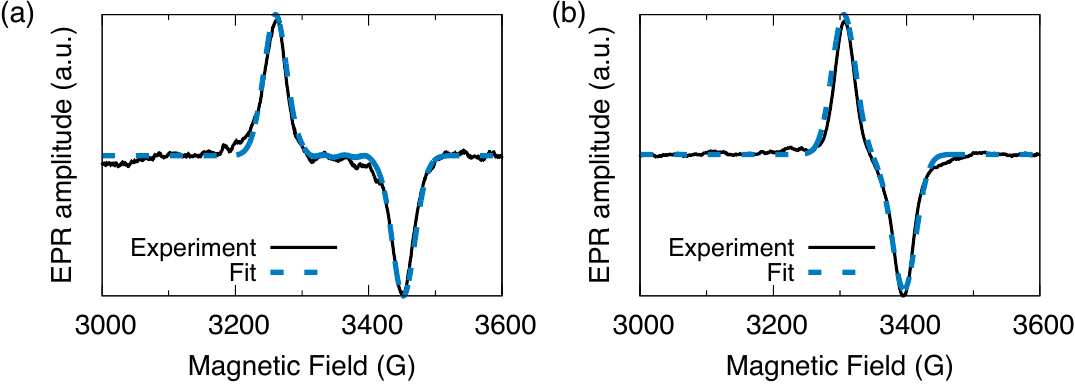}
\caption{
Room temperature EPR spectrum (black solid lines) of AlN in the dark with the magnetic field (a) parallel
and (b) perpendicular to the $c$-axis of AlN along with a fit to the spectra (blue dashed lines).
}
\label{fig:epr}%
\end{figure}

The EPR spectra of sample B obtained at room temperature in the dark are shown in Figure \ref{fig:epr}.
Simulations of the spectra obtained between 300 K and 4 K 
reveal that the HF parameters have a
weak linear dependence with increasing temperature.  Our measured HF parameters are similar to
the prior EPR measurements reported for the D5 spectrum \cite{soltamov2010identification,mason1999optically,evans2006electron}.
Accurate measurement of the $g$-factor in the presence of a strong hyperfine as seen for D5 
is best done by using two different frequencies as was done by 
Soltamov {\it et al.} \cite{soltamov2010identification} where they found $g_{\parallel}$=2.002 and $g_{\perp}$=2.006.
Our values obtained at 10 GHz are $g_{\parallel}$=2.001(1) and $g_{\perp}$=2.004(1),
which is in reasonable agreement with prior measurements.

The HF parameters obtained from our measurements are listed in Table \ref{tab:HF},
along with prior experimental reports for the D5 defect. ENDOR and ENDOR-induced EPR
performed by others have convincingly determined that the defect has axial symmetry
about the $c$-axis, and that the defect is situated on a nitrogen site \cite{evans2006electron}.
\begin{table}[]
\begin{center}
\caption{
Hyperfine parameters that are parallel (A$_{\parallel}$) and perpendicular (A$_{\perp}$) to the $c$-axis,
compared with experimental parameters reported in prior studies of AlN.  For Sample B the number in [ ] is the hyperfine
parameter measured at 4 K.  For the three experimental references the measurement temperature is listed in ( ).
The final three rows contain our first-principles calculations of the HF parameters.
For each calculation, we report the magnitude of the 
HF parameters for the four Al ions that are nearest-neighbor to the defect site where the
fourth entry is the HF parameter of the axial Al ion.
}
\setlength{\tabcolsep}{6pt} 
\renewcommand{\arraystretch}{1.4} 
\begin{tabular}{l cc}
  \toprule\toprule
   & {A$_{\parallel}$ (mT)} & {A$_{\perp}$ (mT)} \\
  \midrule
  \multicolumn{3}{c}{Experiment} \\
  \midrule
  Sample B 300 K [4 K] & 3.2 [4.0] & 1.5 [1.8] \\
  Ref.~\onlinecite{evans2006electron} (40 K) & 4.0 & 1.9 \\
  Ref.~\onlinecite{mason1999optically} (1.7 K) & 3.7 & 1.5 \\
  Ref.~\onlinecite{soltamov2010identification} (9 K) & 4.0 & 1.9 \\
  \midrule
  \multicolumn{3}{c}{First-principles calculations} \\
  \midrule
  O$_{\rm N}$ & 0.9, 0.9, 20.6, 0.4 & 0.6, 0.6, 7.2, 0.2 \\
  $V_{\rm N}$ & 8.8, 4.6, 5.5, 8.6 &  7.0, 3.1, 4.2, 6.7 \\
  C$_{\rm N}$ & 0.3, 0.3, 0.3, 4.0 & 0.7, 0.7, 0.7, 1.7 \\
   \bottomrule\bottomrule
\end{tabular}
\label{tab:HF}
\end{center}
\end{table}

Calculated HF parameters for the possible candidates for D5 are also listed in Table \ref{tab:HF}.
We consider $V_{\rm N}$ and O$_{\rm N}$, since they have been reported as possible candidates in prior
EPR studies \cite{soltamov2010identification,mason1999optically,evans2006electron}.  We also consider C$_{\rm N}$, since
our SIMS measurements indicate carbon is unintentionally incorporated (Table \ref{tab:sims}) in the AlN substrates. Prior first-principles
calculations \cite{lyons2014effects} have also shown C$_{\rm N}$ has low formation energies compared to either C$_{\rm Al}$ or incorporation as an interstitial.

Our calculations provide additional evidence to rule out $V_{\rm N}$ and O$_{\rm N}$ as the origin of the D5 defect.
With O$_{\rm N}$, we find the HF on the axial Al is an order of magnitude lower than the experimental value and the largest
HF associated with O$_{\rm N}$ is on a basal plane Al ion, which is in contrast with the axial anisotropy that is
associated with D5. The HF parameters for $V_{\rm N}$ indicate HF interaction with all four nearest-neighbor Al dangling bonds, which also conflicts with the observed anisotropy of D5. In contrast, the calculated HF parameters for C$_{\rm N}$ and the out-of-plane anisotropy is in remarkably good agreement with the experimental HF parameters listed in Table \ref{tab:HF}.

In the EPR active state (C$_{\rm N}^{0}$), the three basal plane C-Al bonds are equivalent in length and shorter than
the axial C-Al bond length.  The spin density of C$_{\rm N}^{0}$ is localized primarily
on a C $p_{z}$-orbital that is oriented along the $c$-axis \cite{lyons2014effects}.
In our HF calculations, we assume C is incorporated
as $^{12}$C (i.e., the zero nuclear spin isotope of C).
While the majority of the spin density is localized on the C ion (which lacks an HF interaction as
$^{12}$C), a striking consequence of the C $p_{z}$-orbital that constitutes the spin density of C$_{\rm N}^{0}$
is the finite orbital overlap of the C $p_{z}$-orbital with the orbitals of the Al ion that is along the axial C-Al bond.
This orbital overlap leads to a finite HF interaction on the axial Al ion and 
the largest of the four HF parameters for C$_{\rm N}^{0}$
reported in Table \ref{tab:HF} that are in agreement with experiment.
There is negligible overlap of the spin density
with the three basal plane C-Al bonds, which is consistent with the HF parameters
for these three Al ions being significantly lower compared to that of the axial Al.
In reality, 1.1\%~of the C is present as $^{13}$C, which has a non-zero nuclear spin.
Given the overall low concentration of $^{13}$C with respect to the total carbon incorporated in our
samples (Table \ref{tab:sims}), we expect no measurable EPR spectum due to contributions from $^{13}$C.

\begin{figure}[!h]
\includegraphics[width=8.5cm]{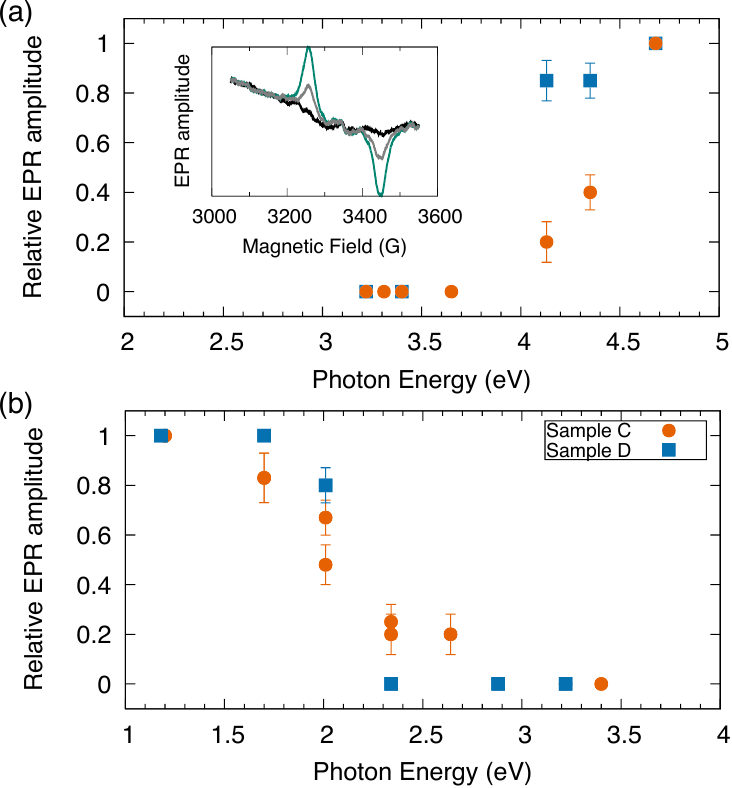}
\caption{
Steady state photo-EPR for (a) excitation and (b) quenching of the EPR amplitude in
 sample C (circles) and sample D (squares) indicating the
relative number of EPR active defects as a function of photon energy.
Error bars indicate the uncertainty in comparing EPR amplitudes.
The inset in (a) illustrates the EPR spectra obtained from
sample C in the dark (black), after illumination with 4.68 eV (green) and following partial quenching
with 2.01 eV (gray)  
}
\label{fig:photoepr}%
\end{figure}

Next we examine possible photoionization processes for C$_{\rm N}$ to determine
the response of the EPR signal to photoexcitation.
Our photo-EPR measurements performed on sample C and sample D
are illustrated in Figure \ref{fig:photoepr}.
The increase in intensity is due to photoionization that leads to C$_{\rm N}^{0}$, while the reduction in intensity
corresponds to conversion of C$_{\rm N}^{0}$ to an EPR silent charge state.
From Fig.~\ref{fig:photoepr}(a), we see that the intensity of the EPR signal begins to increase
at a threshold of $\sim$ 4 eV. Fig.~\ref{fig:photoepr}(b) shows that once the EPR signal is generated,
the intensity begins to decrease at a threshold of $\sim$ 2 eV.

The thresholds for excitation and quenching of the photo-EPR signal add approximately to the band gap of AlN. This suggests the photo-EPR data is due to direct
photoionization of C$_{\rm N}$, rather than the capture of a carrier by a secondary defect followed by charge transfer to C$_{\rm N}$.
This interpretation is further supported by the fact that the photo-EPR response is obtained on two
samples that were grown under different conditions and contain impurities that differ in concentration by 
up to two orders of magnitude.
Finally we note that, no other EPR active defect was observed in sample C,
and no other defects with the appropriate photo-EPR response were detected in sample D.

\begin{figure}[!h]
\includegraphics[width=8.5cm]{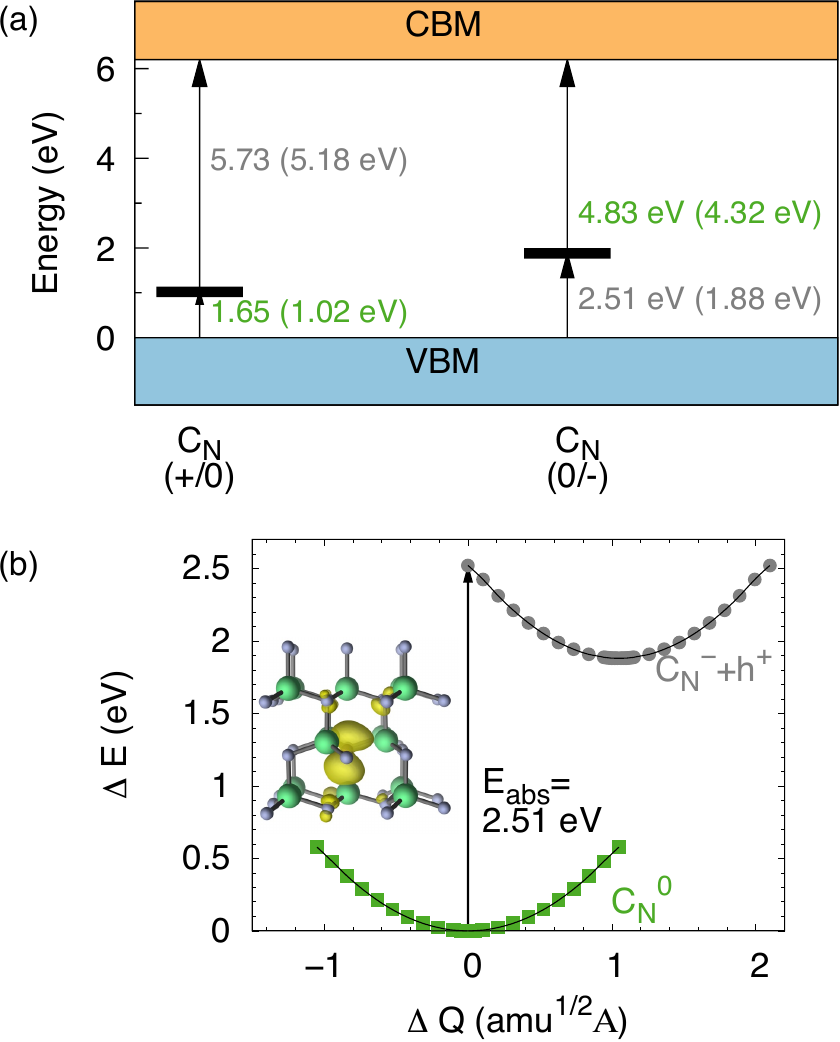}
\caption{
The peak absorption energy
and zero-phonon line (in parentheses) for the
C$_{\rm N}$ (+/0) and (0/$-$) levels is listed alongside each arrow denoting possible
optical absorption processes.
Processes that lead to an EPR active state are denoted in green font.
(b)  Configuration-coordinate diagram illustrating
photoionization of electrons from the VBM to the C$_{\rm N}$ (0/$-$) level converting the EPR active
charge state to an EPR silent state.
The inset in (b) illustrates the spin density of the hole introduced by the EPR active C$_{\rm N}^{0}$ state.
}
\label{fig:dft}%
\end{figure}

First-principles calculations show that in addition to the (0/$-$), which is 1.88 eV above the AlN VBM, C$_{\rm N}$
can also be stable in the positive charge state. The resulting (+/0) level of C$_{\rm N}$ is 1.02 eV above the VBM \cite{lyons2014effects}.
In principle, photoionization processes involving either the (0/$-$) or (+/0) level levels can excite
or quench the EPR-active C$_{\rm N}^{0}$ state.
Our first-principles calculations of the optical transitions
suggest that the photo-EPR measurements correspond to photoionization of the C$_{\rm N}$ (0/$-$) level.
In Figure \ref{fig:dft}(a), we illustrate the zero-phonon line and the peak optical absorption
energies for photoionization processes that involve either the (+/0) and (0/$-$) level
of C$_{\rm N}$ in AlN.  The peak absorption energies are determined by invoking
the Franck-Condon approximation, which is illustrated
using configuration coordinate diagrams as shown in Figure \ref{fig:dft}(b)

Figure \ref{fig:dft}(b) illustrates the photoionization of a carrier
from the AlN VBM that converts the EPR active C$_{\rm N}^{0}$ to C$_{\rm N}^{-}$. This transition is calculated to occur at a peak absorption energy of 2.51 eV, while the onset in the absorption process is 1.88 eV. This prediction agrees with our observation of the photo-EPR signal quenching at $\sim$2 eV.
Similarly, we predict that photoionization of an electron to the AlN CBM from the (EPR silent) C$_{\rm N}^{-}$
charge state occurs with a peak absorption energy of 4.83 eV and an onset of 4.32 eV.
Again, this prediction agrees well with our observation of the photo-excitation of the EPR signal at $\sim$4 eV.
In addition, we find that the photo-EPR cannot involve the (+/0) level of C$_{\rm N}$, since the energies that
excite and quench the EPR signal in our experiments are incompatible with the corresponding energies that we predict from first principles.

Having established evidence for C$_{\rm N}$ in AlN, we are now in a position to comment on optical experiments where carbon has been invoked as the
origin of defect-related spectra.  In particular, several studies have connected the presence of absorption bands at 2.6 eV
and 4.7 eV as being due to carbon \cite{gamov2021photochromism, collazo2012origin,alden2018point, irmscher2013identification},
and there is some debate as to whether carbon incorporates as C$_{\rm N}$ \cite{alden2018point}, 
C$_{\rm Al}$ \cite{gamov2021photochromism}
or as a complex \cite{irmscher2013identification}. 
Given that the threshold to excite the photo-EPR signal due to 
C$_{\rm N}^{0}$ is 4 eV and the photo-EPR signal continues to increase up to 4.68 eV, we suggest that 
the widely observed absorption band at 4.7 eV in AlN can be explained 
by photoionization of C$_{\rm N}^{-}$, leading to an electron in the AlN CBM.

In summary, photo-EPR measurements and first-principles calculations provide clear evidence of carbon incorporation on the nitrogen site in AlN.
The distortion of C$_{\rm N}$ along the $c$-axis leads to a finite hyperfine interaction with the axial Al ion that is nearest neighbor to C$_{\rm N}$.  This resolves the origin of the EPR spectra in a number of prior studies on AlN where the defect giving rise to the hyperfine interaction was thought to be a deep donor. Photoionization of the C$_{\rm N}$ (0/$-$) level at $\sim$4 eV excites the EPR signal associated with this hyperfine interaction, while photon energies in the range of $\sim$2 eV quench the EPR signal, which is consistent with our calculations of the zero-phonon line and peak absorption energies for the C$_{\rm N}$ (0/$-$) level.  The photo-EPR signal excited above 
4 eV due to C$_{\rm N}$ is also consistent with the absorption spectrum that is peaked at 4.7 eV in AlN samples that contain carbon.
\\
\begin{acknowledgements}
We thank Prof. Larry Halliburton for helpful discussions regarding the EPR data.
D.W and J.L.L were supported by the Office of Naval Research through the Naval Research Laboratory's Basic Research Program.  The calculations were supported in part by the DoD Major Shared Resource Center at AFRL.
The work at UAB was supported as part of the Ultra Materials for a Resilient Energy Grid, an Energy Frontier Research Center funded by the U.S. Department of Energy, Office of Science, Basic Energy Sciences under Award No. DE-SC0021230.
\end{acknowledgements}

%

\end{document}